\documentclass[journal]{IEEEtran}

\usepackage{graphicx}
\usepackage{booktabs}
\usepackage{wrapfig}
\usepackage{blindtext}
\usepackage{cite}
\usepackage{hyperref}
\usepackage{url}
\usepackage[center]{caption}
\usepackage[dvipsnames]{xcolor}
\usepackage[final]{changes}
\usepackage{amssymb}
\usepackage{pifont}
\newcommand{\xmark}{\ding{55}}%
\usepackage{colortbl}

\begin{document}

\title{Automating Internet of Things Network Traffic Collection with Robotic Arm Interactions}
%
%

\author{Xi~Jiang$^1$
        and Noah~Apthorpe$^2$\\
         \begin{small} $^1$Department of Computer Science, University of Chicago, Chicago, IL, 60637, USA\\%
    $^2$Department of Computer Science, Colgate University, Hamilton, NY, 13346, USA\\%
    Email: xijiang9@uchicago.edu; napthorpe@colgate.edu
    \end{small}
}

%
%

\markboth{}%
{Jiang \MakeLowercase{\textit{et al.}}: Automating Internet of Things Network Traffic Collection with Robotic Arm Interactions}

%



\maketitle

\begin{abstract}
Consumer Internet of things research often involves collecting network traffic sent or received by IoT devices. These data are typically collected via crowdsourcing or while researchers manually interact with IoT devices in a laboratory setting. However, manual interactions and crowdsourcing are often tedious, expensive, inaccurate, or do not provide comprehensive coverage of possible IoT device behaviors. 
We present a new method for generating IoT network traffic using a robotic arm to automate user interactions with devices. 
This eliminates manual button pressing and enables permutation-based interaction sequences that rigorously explore the range of possible device behaviors.
We test this approach with an Arduino-controlled robotic arm, a smart speaker, and a smart thermostat, using machine learning to demonstrate that collected network traffic contains information about device interactions that could be useful for network, security, or privacy analyses.
We also provide source code and documentation allowing researchers to easily automate IoT device interactions and network traffic collection in future studies. 

\end{abstract}


%
\IEEEpeerreviewmaketitle

\section{Introduction}
\label{sec:introduction}
The Internet of things (IoT) refers to the wide variety of physical objects increasingly connected to the Internet. IoT devices range from common household items, such as thermostats, light bulbs, and door locks, to medical products, wearables, and industrial sensors. For example, a ``smart'' (IoT) thermostat may be able to receive location data from a smart car and automatically adjust the home temperature when a user leaves work.
Although IoT devices provide unprecedented convenience and efficiency, they also raise concerns about security and privacy. Users report privacy fears associated with constant connectivity and always-on environmental sensors~\cite{naeini2017privacy, zheng2018user, apthorpe2020you, huang2020amazon}, and researchers have identified insecurities in IoT device software and network communications~\cite{loi2017systematically, chu2018security, wang2019looking}. These issues, combined with the increasing popularity of consumer IoT devices, are motivating many studies analyzing IoT network traffic to detect and prevent privacy and security vulnerabilities. 

Collecting IoT network traffic for these studies typically involves researchers manually interacting with devices in a laboratory setting. Researchers press buttons, touchscreens, or other user interface elements on the devices in an attempt to mimic real-world user behaviors or collect traffic from as many device states as possible. For example, a researcher may attempt to record network traffic while pressing every button on a device in sequence or by pressing patterns of buttons (or other user interface elements) in order to examine the full scope of device behavior~\cite{apthorpe2019keeping}. However, these types of manual device interactions are time-consuming and are unlikely comprehensive, posing a significant challenge to IoT network research (Section~\ref{sec:related}). 

IoT studies have also utilized crowdsourcing to acquire consumer IoT network traffic~\cite{huang2020iot, mazhar2020characterizing}. This can provide large quantities of realistic data; however, crowdsourcing typically requires extensive data collection platform development, and user reports of device behavior during traffic collection are prone to inaccuracy. 

In this paper, we present a novel method for automating IoT network traffic collection: configuring a robotic arm to interact with IoT devices according to formalized interaction sequences (Section~\ref{sec:method}), particularly on IoT devices with physical user interface (UI) elements such as buttons and switches. We focus on devices with physical UI elements because the collection of network traffic on such devices has been proven to be challenging and costly (Section \ref{sec:related}). At the same time, the application of robotic arm-assisted traffic collection is more suitable for these devices compared to devices with non-physical UI elements, which we discuss in Section \ref{sec:limit}.
This approach can simulate real user behaviors, provide comprehensive coverage of possible user/device interactions, and eliminate tedious manual button pressing. Recording Internet traffic to and from a device during robotic arm interactions provides a source of network data that can be used for security testing, privacy evaluation, or other IoT research.
As far as we are aware, there has been no prior use of robotics to automate IoT research in this manner, raising significant potential for cross-disciplinary follow-up research.

We demonstrate the effectiveness of our approach by configuring a robotic arm to press physical buttons on two devices: an Amazon Echo Show 5~\cite{echo} (a smart speaker with multiple user interfaces) and an Emerson Sensi Wi-Fi Smart Thermostat~\cite{sensi} (a household thermostat with only physical buttons) (Section~\ref{sec:evaluation}). We verify that this produces network behavior correlated with robot/device interactions by testing a variety of permutation-based button press sequences while collecting Internet traffic. We then train a machine learning model (random forest classifier) to accurately infer ($F_1 > 0.95$) which specific buttons are pressed on the devices from network traffic alone. 
This machine learning task is inspired by the various metadata-based inference attacks in the consumer IoT privacy literature~\cite{apthorpe2019keeping, acar2020peek, trimananda2020packet}.
Success at this task indicates that the captured traffic provides substantial information about interactions with the devices 
and corresponding device behavior and would be useful for follow-up network, security, or privacy analyses.

Employing robotics for collecting network data from physical IoT devices eliminates many drawbacks of manual data collection or crowdsourcing. Our approach provides rigorous interaction coverage and high scalability, allowing for easier collection of Internet traffic for IoT network audits. 
Although we focus on network data collection in this paper, there is great potential for robotic automation of other research involving cyber-physical devices, including fuzz testing and usability testing.
We hope that this approach and our provided source code\footnote{\href{https://github.com/Chasexj/Automated\_IoT\_Traffic\_Generation}{https://github.com/Chasexj/Automated\_IoT\_Traffic\_Generation}} will facilitate continued IoT research (Sections~\mbox{\ref{sec:limit}--\ref{sec:future}}).
\section{Background \& Related Work}
\label{sec:related}

Collecting network data from consumer IoT devices has posed a consistent challenge for IoT research. Studies typically use either manual data collection or crowdsourcing to collect IoT network traffic; however, both approaches have substantial drawbacks. 
In this section, we review the challenges of these traditional methods for IoT data collection and discuss how our approach provides significant advantages in terms of automation, monetary cost, verifiability, and scalability (Table~\ref{tab:compare}).

\begin{table}[t]
\centering\footnotesize
\begin{tabular}{lllll}
\toprule
& \textbf{Manual}  & \textbf{Crowdsourcing}  & \textbf{Robotic Arm}       \\ \midrule
\textbf{Automated}          & \textcolor{red}{\xmark}  & \textcolor{red}{\xmark}   & \textcolor{green}{\checkmark}        \\
\textbf{Cost}         & \textcolor{red}{\$\$}     & \textcolor{red}{\$\$\$}          & \textcolor{green}{\$}      \\ 
\textbf{Verifiable} & \textcolor{green}{\checkmark}      & \textcolor{red}{\xmark}           & \textcolor{green}{\checkmark}     \\ 
\textbf{Scalable}      & \textcolor{red}{\xmark}    & \textcolor{green}{\checkmark}         & \textcolor{green}{\checkmark}       \\ \bottomrule
\end{tabular}
\caption{Comparison of research methods for IoT traffic collection. Our robotic arm method automates verifiable physical interactions with IoT devices without extensive manual effort or expensive crowdsourcing campaigns.}
\label{tab:compare}
\end{table}     

\subsection{Manual IoT Traffic Collection}
Many studies of consumer IoT network traffic have involved data collection in a controlled laboratory environment. 
Researchers acquire devices relevant to their research question, instrument the devices,
and then manually interact with the devices to test their behavior and collect data for online and offline analysis.
Studies utilizing manual data collection have focused on privacy-violating inferences from IoT network traffic~\cite{apthorpe2019keeping, acar2020peek, edu2020smart} and network vulnerabilities in specific classes of devices (e.g., children's toys~\cite{chu2018security, shasha2019playing}), sparking increased consumer awareness, regulatory action, and manufacturer attention to consumer IoT security and privacy. 

Manual data collection has several benefits. First, researchers can precisely control the network environment, ensuring that recorded network traffic actually corresponds to specific devices or user interactions. This greatly simplifies ground-truth labeling of collected network traffic for input into supervised machine learning algorithms or for other follow-up analyses.
Second, researchers can collect data while subjecting the device or network to active attacks that would be unethical outside of a controlled environment. For example, researchers can flood a device with denial-of-service traffic or attempt to install bogus TLS certificates. 

Unfortunately, manual data collection also has several serious drawbacks that have limited consumer IoT research. First, collecting network traffic from all possible user interactions and device behaviors is usually infeasible. 
Most consumer IoT devices do not have emulator support, so user interactions must be tested on a physical device.
Manually testing repeated sequences of specific user interactions (e.g., button presses) on a device is quite tedious, limiting data collection to a small set of interactions and a correspondingly small volume of network traffic. 
In contrast, our robotic arm approach does not require any manual interaction or intervention during data collection. This can save hours of researcher time and eliminate the possible need to restart experiments if a human researcher forgets a button press or presses the wrong button. 

Second, manual researcher interactions with consumer IoT devices are unlikely representative of real user behavior over varying timescales. This means that network traffic generated via manual interactions may not be representative of traffic generated by devices in consumer households. In contrast, our approach uses permutation-based interaction sequences that can test all possible interactions with a device or be programmed to mimic real user behavior over long timescales.

\subsection{Crowdsourcing}
A more recent approach to acquiring consumer IoT network traffic involves crowdsourcing data collection to real users who have adopted IoT devices~\cite{huang2020iot, mazhar2020characterizing}. This typically involves the creation of custom hardware or software that allows users to instrument their own devices or home networks.

Crowdsourcing IoT traffic has several benefits. Crowdsourced data is more externally valid than manually collected laboratory data because it comes from real users interacting with IoT devices in the wild. 
With substantial recruitment efforts, crowdsourcing can also produce more data from a wider variety of devices and interaction patterns than laboratory collection. 

However, crowdsourced data has other drawbacks. It may require extensive development effort to create a platform for data collection that participants feel comfortable incorporating into their homes. Recruiting participants is also challenging. Crowdsourcing platforms such as Amazon Mechanical Turk are not well suited to IoT data collection, which does not fit well into the Human Intelligence Task (HIT) framework.
Participant compensation for large-scale crowdsourcing campaigns can also be quite expensive, especially if participants are required to install hardware in their homes. 
Some studies avoid this expense by asking interested individuals to volunteer data without monetary compensation, sometimes by providing details about local IoT device behavior that may be of interest to privacy or security-conscious users~\cite{huang2020iot}. However, this can limit study participation to technically savvy participants or those with existing privacy or security concerns, potentially introducing results bias. Our robotic arm approach requires only an initial cost for the robot and devices but, like crowdsourcing, can produce large amounts of data.

Crowdsourced IoT data also suffers from unreliable labels. Participants may not accurately report what devices they own or what interactions they perform with the devices. \textit{Post hoc} identification and verification of device types and user interactions from network data is a research problem in its own right~\cite{huang2020iot}.
In contrast, data collected with our automated approach can be directly labeled with the correct ground truth by the researchers running the robot platform.

\subsection{Automated Data Collection with Robotic Arm Interactions}
The automated data collection approach described in this paper combines the scalability and reduced tediousness of crowdsourcing with the control and verifiability of manual data collection (Table~\ref{tab:compare}). 
Our approach focuses on automating device interactions via \textit{physical} user interface elements rather than through networked applications (e.g., smartphones, IoT hubs, or debugging tools). While tools like Android Debug Bridge (ADB) can be used to automate interactions with some IoT devices~\cite{ren2016recon,mohajeri2019watching}, many devices do not have similar programmatic testing tools available. Such devices can only be tested with physical interactions (e.g., button pressing). By performing these physical interactions with a robotic arm, we are able to examine network traffic from a breadth of device behaviors with minimal manual effort. 
We are also able to rigorously test permutations of device interactions to collect data for all possible (or all reasonable) interactions with a device. These permutations can include different interface elements (buttons, etc.) as well as different timings between interactions.
\section{Method}
\label{sec:method}

Our approach for generating automated IoT device traffic has two main components: 1) Configuring a robotic arm to rigorously test user interface interactions with IoT devices, and 2) Automating network traffic collection during robotic arm interactions via existing IoT traffic analysis tools~\cite{iot_inspector}.

Implementing this approach involves four primary challenges: 1) Efficiently obtaining correct input parameters for the robotic arm such that it interacts with desired user interface elements on the IoT device, 2) Ensuring interaction accuracy between the robotic arm and the IoT device, 3) Designing interaction sequences that thoroughly explore the space of possible device behaviors, and 4) Automating network traffic collection during robot/device interaction. 

This section describes our implementation, including how we addressed these challenges for the Amazon Echo Show 5 and Sensi Wi-Fi Smart Thermostat devices. Our source code is publicly available for use in future research.

\subsection{Experiment Setup}
\label{sec:experiment-setup}
\subsubsection{Robotic Arm} 
\begin{figure}
    \centering
    \includegraphics[width=8.5cm]{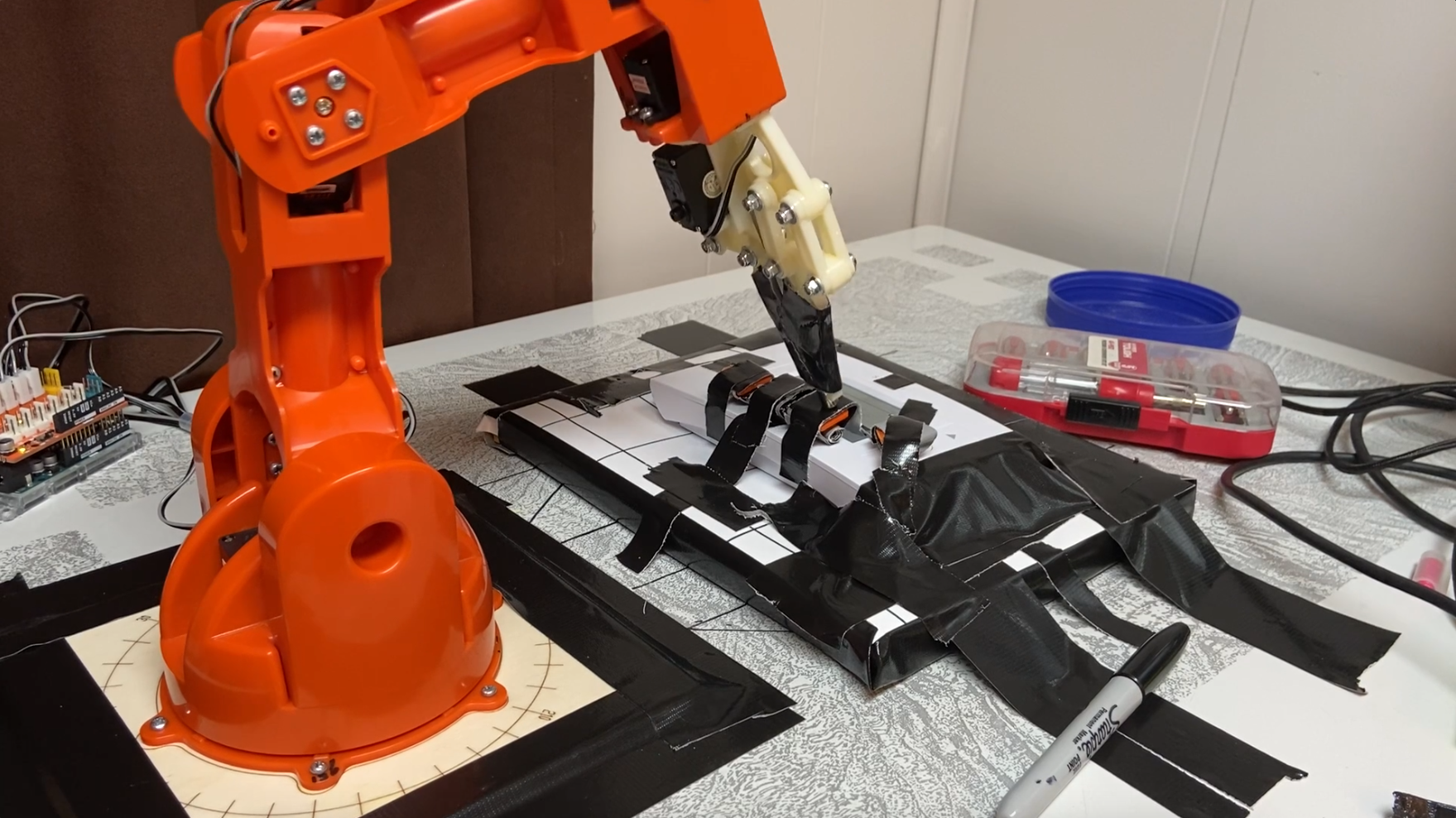}
    \caption{Sample hardware setup with Arduino Braccio Robotic Arm \textit{(left)} and IoT device \textit{(right)} fixed to the table.}
    \label{fig:set_up}
\end{figure}

The main robotic arm used in this study is the Arduino Braccio Robotic Arm~\cite{braccio} with 6 degrees of freedom (DoF) and motors/servos connected to the Braccio shield. The robotic arm is fixed to the work station as illustrated in Figure~\ref{fig:set_up} and is controlled via an Arduino UNO R3 board~\cite{r3}. 
When given correct rotation and movement delay parameters, the robotic arm can press buttons and interact with other interface elements on an IoT device fixed within the arm's maximum reach.

We chose this particular model of robotic arm for two main reasons: 1) Unlike industrial robotic arms which often sell for tens of thousands of dollars, the Arduino Braccio Robotic Arm retails for less than \$250. This keeps our method accessible for IoT researchers with limited budgets, 2) Compared to other robotic arms in the same price range, this model has a relatively high precision ($\pm2$mm), which is desirable for interacting with devices with small buttons.

\subsubsection{IoT Devices}The two devices tested in this study are the \textbf{Amazon Echo Show 5 (Echo-Show5)}~\cite{echo} and the \textbf{Emerson Sensi Wi-Fi Smart Thermostat (Sensi-Thermostat)}~\cite{sensi}. Echo-Show5 is an Amazon smart display capable of video and voice calls, content streaming, online shopping, and running third-party applications. The device is equipped with Amazon's Alexa~\cite{alexa} voice assistant, a touch screen, as well as four buttons on the top of the device. The four buttons include a mechanical switch for the camera shutter, a microphone mute button, and two buttons responsible for volume up/down. Sensi-Thermostat is a smart Wi-Fi thermostat that is designed to control conventional household heating/cooling. While it can be connected to a user's smartphone for remote control, there are also six physical buttons on the device that can be used to raise/lower temperature, turn on/off the fan, and change the heating/cooling mode and schedule.

We chose these specific IoT devices based on two specific criteria:

\paragraph{Market representativeness} While there are numerous IoT devices on the market, we wanted to select representative devices that are widely deployed to evaluate the practicality
and scalability of our automated data collection method. We chose the Echo-Show5 because it is representative of the IoT smart speaker/display market (Amazon is a major IoT manufacturer with over 51\% of the U.S. smart speaker market share as of 2020~\cite{amazon_data}). Likewise, the Sensi-Thermostat is a popular IoT device with $>$16,000 Amazon reviews circa January 2022, making it one of the most widely deployed IoT smart thermostats on the market. Although there are many categories of IoT devices that we are not testing in this study, our methodology is highly transferable
across different devices as long as physical user interface (UI) elements are present. We also believe that the approach can be adapted to additional devices with
a wide variety of user interfaces (Section \ref{sec:limit}).

\paragraph{Network behavior complexity} It is essential for us to verify that our proposed method is effective across devices with different levels of network behavior complexity. 
The Echo-Show5 is a comparatively sophisticated device that offers a significantly wider variety of functionalities than single-purpose IoT devices such as smart light bulbs or outlets, allowing us to evaluate our approach on a device with complex network behaviors. On the other hand, the Sensi-Thermostat is a single-purpose device, making it ideal for validating our automated data collection approach for IoT products with simpler network behavior.

\subsubsection{Network Traffic Collection} 
\label{sec:network-collection}
To capture network traffic to and from the devices, we set up a Raspberry Pi Wi-Fi access point using existing ``IoT Inspector'' software~\cite{iot_inspector}. This configures the Raspberry Pi such that all traffic through the Wi-Fi network is captured and stored locally as PCAP files. Upon setting up the robotic arm, IoT device, and Raspberry Pi access point, we instruct the arm to interact with the device as described in the following section. These interactions cause the device to generate bidirectional network traffic that is recorded for analysis.
Since the recorded PCAP files contain the source and destination addresses of captured network packets, we can use tools such as Wireshark~\cite{wireshark} to separate packets generated specifically by the IoT device from background traffic.
To test this approach with the Echo-Show5 and Sensi-Thermostat, we filter the captured traffic to include only 
Transmission Control Protocol (TCP) packets sent to or from these devices' Ethernet MAC addresses. We also include background TCP traffic collected during periods without device interactions (idle time) to help evaluate the effectiveness of our approach.

\subsection{Inverse Kinematics for Robotic Arm Input Parameters}
\label{sec:inverse-kinematics}
Once the robotic arm and IoT devices are fixed in place, we calculate the arm joint rotation parameters needed for the arm to reach the buttons on the device. 
For each button, we manually measure the base rotation needed to align the arm with the button followed by the button's 2D vertical coordinates with respect to the base of the arm. 

We then use the inverse kinematics scripts from~\cite{invkin}, which take the lengths of each section of the arm and each button's 2D coordinates and output the needed arm rotation parameters. 
These scripts only apply to robotic arms with 3 degrees of freedom (DoF), so we treat the 6 DoF Arduino Braccio Robotic Arm
as having 4 DoF: one base joint that can rotate horizontally and three arm joints that can rotate vertically.

\subsection{Ensuring Interaction Accuracy}
The buttons on most IoT devices are designed for precise interactions with human fingers and are often relatively small (usually less than 1 cm$^2$). This can be problematic, as the robotic arm can experience small deviations ($\pm$2 mm) during every movement, causing missed presses or unintended presses of adjacent buttons if not corrected.

\begin{figure}
    \centering
    \includegraphics[width=8.5cm]{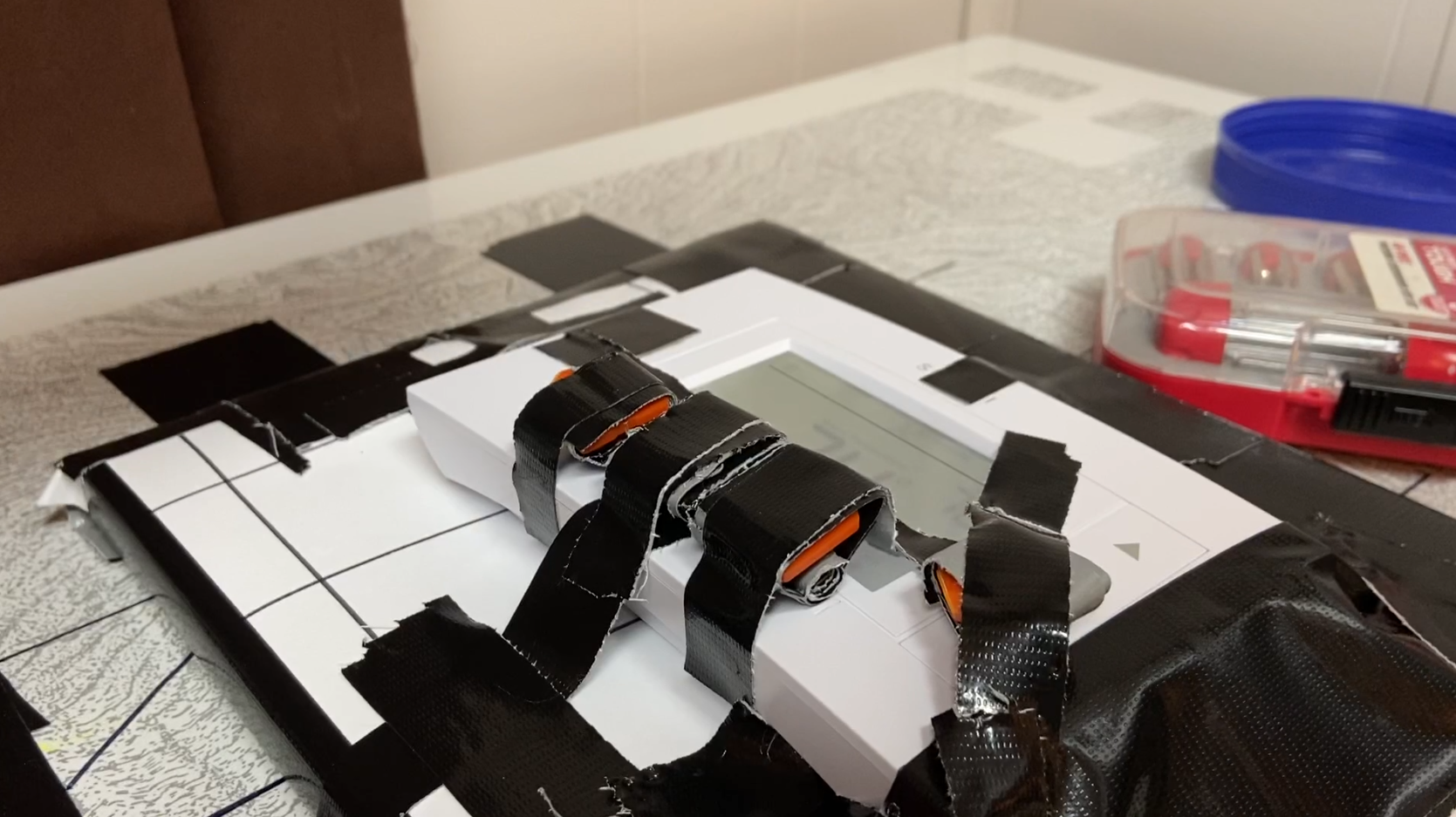}
    \caption{IoT device (Sensi-Thermostat) with physically enlarged buttons to eliminate robotic arm misalignment errors.}
    \label{fig:buttons}
\end{figure}

In our specific case, we address this problem by physically enlarging the buttons' surface areas to overcome the possible deviations from each arm rotation. This is achieved by attaching additional hard surfaces on top of the original buttons (Figure~\ref{fig:buttons}). Given the larger surface areas, such rotation biases become insignificant, and missed presses are avoided.
Our solution is viable because the robotic arm used in the experiment shows randomized movement biases and does not introduce significant accumulated misalignment over the course of long experiments.
However, in scenarios where such deviations indeed accumulate, there are existing tools~\cite{vision-based-robotic-grasping,mariottini2011active,kumar2018computer} that rely on computer vision to perform automated correction and can be applied to address the issue.

While we address the movement bias, it is worth noting that the robotic arm also experiences delays in executing movement commands: when a movement is requested via the Arduino IDE command execution, a very small delay occurs before the robotic arm motors actually move to adjust the arm to the desired position. However, this delay in execution time is on the order of milliseconds and, unlike the positional bias, it is fixed and constant for each movement. This means that the time delay between button press commands and actual button presses is predictable, does not accumulate over multiple commands, and is insignificant compared to the timescale of the entire UI interaction sequences. 

\subsection{Designing Interaction Sequences}
\label{sec:interaction-seqs}
Our system instructs the robotic arm to move to specific locations by providing a set of arm rotations as the input parameters (Section~\ref{sec:inverse-kinematics}). Given the limited number of unique physical buttons on each IoT device, we can configure the robotic arm to iterate through buttons in a specific order, or ``interaction sequence,'' by feeding in consecutive sets of arm rotations.

Our system performs comprehensive device interaction testing through permutation-based interaction sequences. We assign each button on the device a unique number, and then the system instructs the robotic arm to press the buttons in all possible permutations of these numbers. 
By default, the system uses a fixed 10-second delay between button presses, but this and other details can be easily customized. For example, a researcher could choose to use permutations with random repetitions, so the robot will press individual buttons multiple times per interaction sequence. A researcher could also randomize the time between button presses to simulate unpredictable user behavior (Section~\ref{sec:future}). 

\subsection{Demonstrating Effectiveness with Machine Learning}
\label{sec:mlmethod}

We evaluate the effectiveness of our approach by verifying that the button presses performed by the robotic arm actually lead to the collection of useful network traffic data. We do this by demonstrating that we can train a machine learning (ML) model to predict which buttons were pressed on the device from the network traffic alone. This allows us to conclude that the captured traffic provides substantial information about interactions with the IoT device and is therefore relevant for follow-up network, security, or privacy analysis. This machine learning task is inspired by the various metadata-based inference attacks in the consumer IoT privacy literature~\cite{apthorpe2019keeping, acar2020peek, trimananda2020packet}, in which network traffic is shown to reveal substantial information about user interactions with their IoT devices.

It is important to emphasize that the design of the ML algorithm is \textit{not} the primary contribution of this study. We use ML simply as a method to verify that our robotic arm interaction approach generates useful IoT traffic. We expect that this automated method will be effective for many research purposes given that it allows rigorous device testing without tedious manual effort. 

We perform the ML evaluation as follows: 1) Label recorded IoT device packets (training data) with the buttons pressed by the robotic arm immediately prior to collection, 2) Train a supervised machine learning model to predict which button presses caused the device to send specific packets in reserved test data.

\subsubsection{Training Data}
We collect training data by instructing the robotic arm to repeatedly press each button on the device while recording the timestamp of each button press. For example, the arm performs 15 presses of button \#1 followed by 15 presses of button \#2, etc. The delay between each button press is a fixed 10 seconds. 

We record the device's network traffic during these interaction sequences as PCAP files as described in Section~\ref{sec:experiment-setup}.
We convert these PCAP files into a standardized CSV format using the nPrint tool~\cite{holland2020nprint}. nPrint encodes each packet as a row in the CSV file with columns containing the following features: packet's source IP address, destination IP address, payload, IPv4 headers, TCP headers, and relative timestamps. We then add a label column with the integer identifier of the button pressed immediately prior to the corresponding packet as determined by comparing packet timestamps and button press timestamps. A single button press can cause network activity consisting of multiple packets, so for each button press with timestamp $t_{button}$, we treat this button as the label for all packets with timestamps $t_{packet}$ within the following range: 
\[t_{button} \leq t_{packet} < (t_{button} + 10s)\] 
This labeling is reasonable because all associated packets are transmitted a few seconds after a button is pressed---well within the 10 second gap between consecutive button presses. 

\subsubsection{Random Forest Classifier}
We train a random forest classifier~\cite{ho1995random} using the scikit-learn library~\cite{scikit-learn} and our collected training data. 
The classifier is designed to predict which button the robot pressed given only the subsequent network packet data. We select this particular ML algorithm in our study because the effectiveness of using random forest classifiers for characterizing network traffic has long been established~\cite{wang2015network,zhang2014robust,jun2007internet}.

We randomly select 25\% of the labeled data to withhold as a test set and use the remaining 75\% as a training set.
We then perform a grid search to choose model hyperparameters evaluated by multi-metric (accuracy, precision, recall) 10-fold cross-validation over the training set. The grid search considers the criterion function (metric for measuring the quality of a split when constructing a decision tree in the random forest), maximum depth of the decision trees in the random forest, minimum number of samples required to split an internal node when constructing each decision tree, and the total number of trees in the random forest.
Specific hyperparameter values tested and selected by the grid search are shown in Table~\ref{tab:grid_parm}.

\begin{table}[t]
\centering\footnotesize
\begin{tabular}{lllll}
\toprule
\textbf{Hyperparameter}          & \textbf{Value 1}  & \textbf{Value 2}   &\textbf{ Value 3}  & \textbf{Selected}       \\ \midrule
\textbf{criterion}          & gini  & entropy   & & entropy        \\
\textbf{max\_depth}         & 20     & 40          & 80  & 40      \\ 
\textbf{min\_sample\_split} & 2      & 5           & 10  & 2      \\ 
\textbf{n\_estimators}      & 200    & 400         & 800 & 800      \\ \bottomrule
\end{tabular}
\caption{Grid search values for random forest hyperparameters and selected options for final model}.
\label{tab:grid_parm}
\end{table}                                                                         

Using the best hyperparameters found by the grid search, we train a classifier on the entire training set and evaluate its precision, recall, and $F_1$ score on the test set. 
To obtain more reliable results, we repeat the above procedures (randomized train/test split, grid search, and classifier testing) 50 times and compute the means and variances of the test scores.
High scores indicate that the classifier can confidently associate recorded packets with the button presses that generated those packets,
implying that the collected network data captures relevant information about user interactions and device behaviors. 
\section{Results}
\label{sec:evaluation}

Our study demonstrates that a robotic arm can be used to automate interactions with IoT devices in order to collect network traffic for research. Testing this approach with the Echo-Show5 and Sensi-Thermostat shows that it can collect IoT traffic that provides rigorous coverage of device behaviors with high correlations between button presses and captured packets.

\subsection{Visual Support for Automated IoT Traffic Collection}
\label{sec:results-visual}
\begin{figure*}[t]
    \centering
    \includegraphics[width=0.65\textwidth]{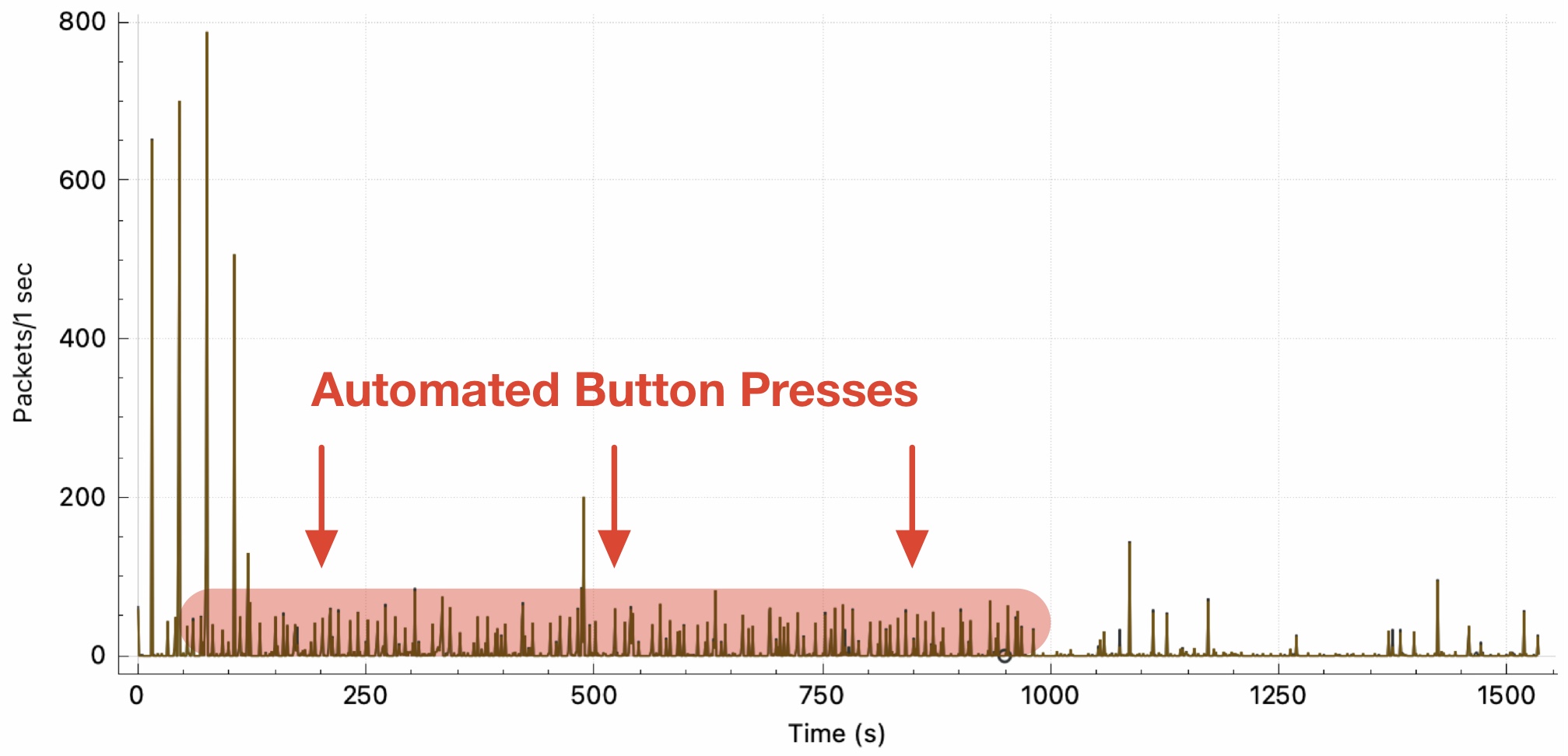}
    \caption{Example Echo-Show5 TCP traffic showing significantly more packets during automated robotic arm interactions. This packet trace was collected over 1533 seconds using a permutation-based interaction sequence (Section~\ref{sec:interaction-seqs}) with 10 seconds between each interaction. The device sent and received 8.83 packets per second on average during the automated button presses and 2.25 packets per second on average after the automated interactions (idle time)}
    \label{fig:echo_io_shade}
\end{figure*}

\begin{figure*}[t]
    \centering
    \includegraphics[width=0.65\textwidth]{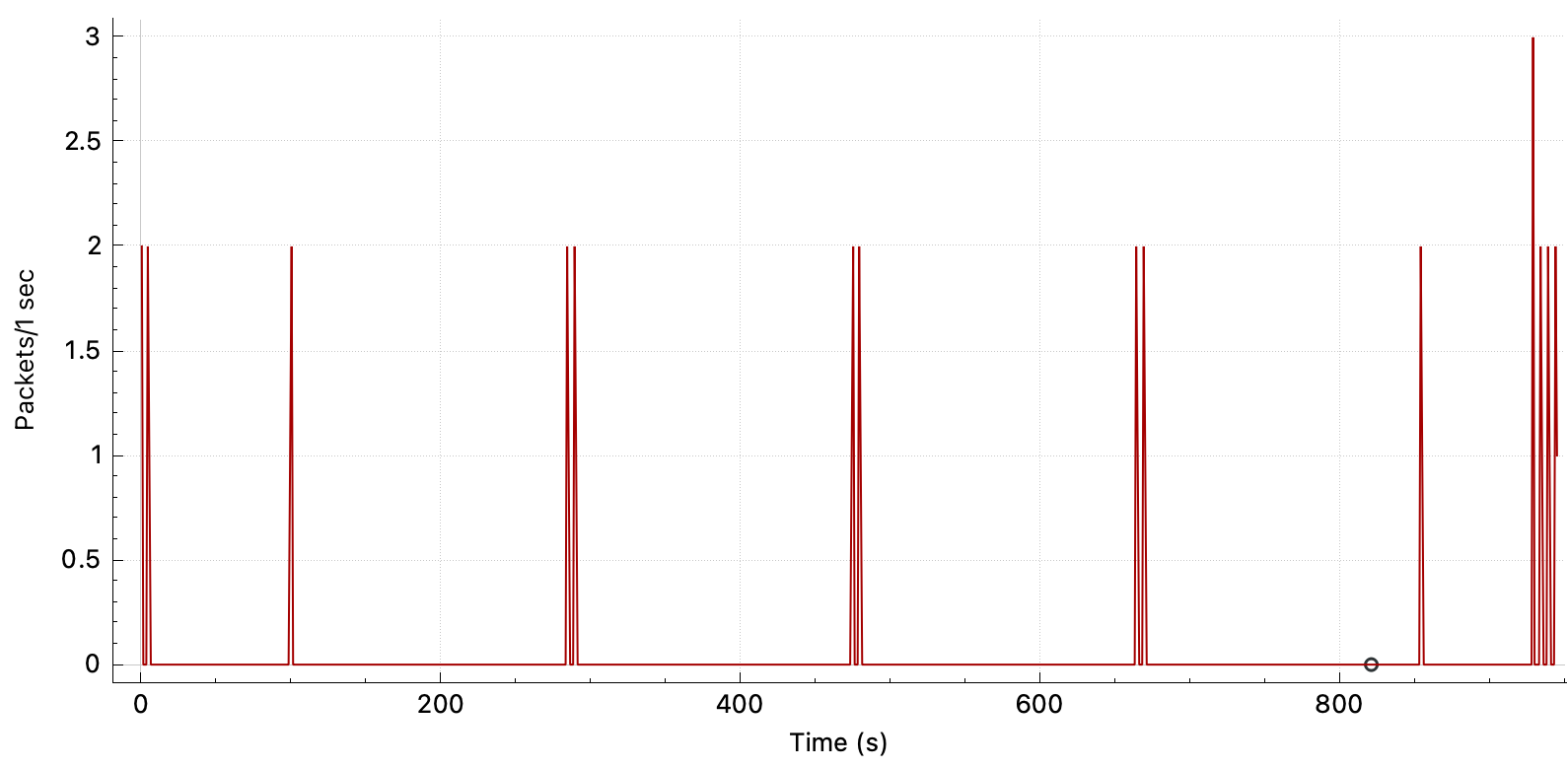}
    \caption{Example Sensi-Thermostat TCP traffic during automated robotic arm interactions. This packet trace was collected over 950 seconds using a permutation-based interaction sequence (Section~\ref{sec:interaction-seqs}) with 10 seconds between each interaction. The device batches periodic messages approximately every 200 seconds in response to settings changes made by the intervening button presses.}    \label{fig:therm_io}
\end{figure*}

Visualizations of traffic traces collected from the Echo-Show5 (Figure~\ref{fig:echo_io_shade}) and the Sensi-Thermostat (Figure~\ref{fig:therm_io}) during automated interactions with all physical buttons on each device verify that our approach produces interaction-correlated network data. We performed both experiments without network congestion to prevent TCP congestion control from causing unwanted variations in the collected traffic. In practice, nothing prevents future applications of our automated approach in network environments with background congestion.

\subsubsection{Echo Show 5} Traffic collection from the Echo-Show5 started at the relative timestamp of 0s and ended at the relative timestamp of 1533s. The robotic arm started pressing buttons on the device at 50s according to a permutation-based interaction sequence and concluded all button presses at 1000s.

There are clear traffic spikes during the time period with the automated button presses. The period with button presses contained a total of 8391 packets with an average of 8.83 packets per second. In contrast, the rest of the packet capture (idle time) contained only 1202 packets and an average of 2.25 packets per second. Note that we excluded packets prior to the beginning of the button presses when calculating these statistics, as these early packets are the result of device initialization at startup. 

The significantly higher amount of network traffic during the button presses indicates that our robotic arm's interactions with the Echo-Show5 indeed generates activity-related traffic compared to periods with no robot interaction. 
These results corroborate Apthorpe et al.'s previous findings~\cite{apthorpe2019keeping} that user interactions with an Amazon voice assistant cause detectable increases in traffic rates.

\subsubsection{Sensi-Thermostat} 

Traffic collection from the Sensi-Thermostat started at the relative timestamp of 0s and ended at the relative timestamp of 950s. The robotic arm button presses occurred from 0s to 950s at intervals of 10s. 

The Sensi-Thermostat sends periodic updates approximately every 200 seconds to reflect changes in settings caused by the intervening button presses. No TCP traffic is exchanged outside of these periodic updates. Unlike the Echo-Show5, the Sensi-Thermostat \textit{does not generate any TCP traffic} when the robotic arm is inactive and no buttons are pressed, because no periodic update is required. This allows us to conclude that the observed traffic spikes at a rate of 2 packets per second occurred as a direct result of the automated button presses. 
Compared to the Echo-Show5, there are significantly fewer captured packets for the Sensi-Thermostat; however, this is expected, as the Sensi-Thermostat is a simpler device.

The intermittently spiking traffic we observe corroborates  Apthorpe et al.'s previous findings~\cite{apthorpe2019keeping} that the traffic patterns of single-purpose consumer IoT devices (e.g., thermostats, lightbulbs, and outlets) are often directly and obviously correlated with user interactions and devoid of substantial TCP background traffic.
The observable correlation between the Sensi-Thermostat's network traffic spikes and the automated robotic button presses validates the effectiveness of our data collection approach for this device. 

\subsection{ML Support for Automated IoT Traffic Collection}
We used machine learning, along with the IoT traffic captured from the Echo-Show5, to verify the correlation between button presses and captured packets as described in Section~\ref{sec:mlmethod}. The success of our machine learning model further supports the ability of our automated interaction method to produce traffic containing information about device behavior that would be useful for network, security, or privacy research.

We performed the machine learning evaluation using the Echo-Show5 data instead of the Sensi-Thermostat data because the Sensi-Thermostat produced such a small amount of traffic (41 packets) that 1) it is easy to visually verify correlations between traffic spikes and button presses (Section~\ref{sec:results-visual}) and 2) there is not enough data to train an ML model. 
In comparison, the Echo-Show5 produced sufficient traffic for random forest training (8391 packets). 

\begin{figure*}[t]
    \centering
    \includegraphics[width=0.7\textwidth]{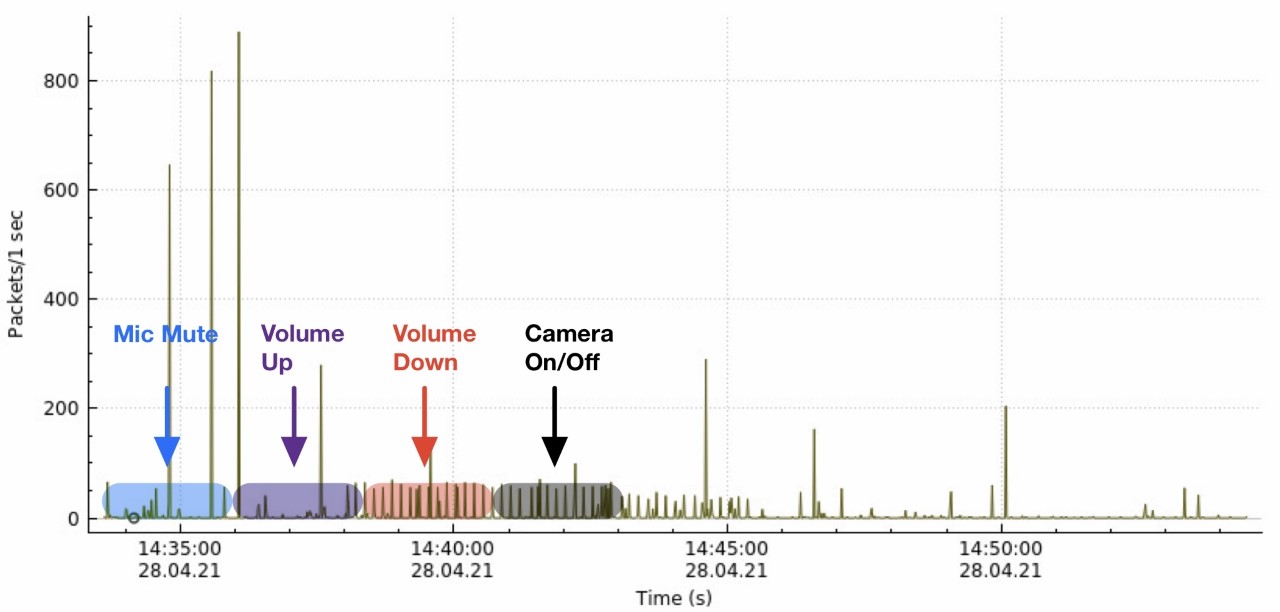}
    \caption{Example Echo-Show5 TCP traffic with labeled button presses. This trace was collected with 15 presses of each of 4 physical buttons on the device and 10 seconds between each button press. The trace contains 8391 total packets.}
    \label{fig:alexa_ml_io}
\end{figure*}

We labeled each packet in the Echo-Show5 traffic with the number of the button most likely responsible for it being sent (Figure~\ref{fig:alexa_ml_io}). We randomly split the labeled traffic into 75\%/25\% training/test sets and conducted a grid search using the training data to choose hyperparameters for the random forest classifier. 
Training the classifier took only a few seconds using the scikit-learn library. 
We then tested the classifier on the test set and recorded the precision, recall, and $F_1$ scores, weighting each button's contribution to the average score by its relative cardinality in the test set. 

We repeated this training and testing process 50 times using the Echo-Show5 traffic as described in Section~\ref{sec:mlmethod}.  All average precision,  recall, and $F_1$ scores were approximately 0.96 with variances less than 1.0e-4, demonstrating a strong correlation between the device's network behavior and the robotic arm button presses (Table~\ref{tab:scores}). 

These high scores allow us to conclude that the captured traffic does provide substantial information about the robot arm interactions and would be useful for follow-up research about the network, security, or privacy implications of user interactions with the IoT device.

\begin{table}[t]
\centering
\begin{tabular}{llll}
\toprule
& \textbf{Precision}   & \textbf{Recall}  & $\mathbf{F_1}$       \\ \midrule
\textbf{Average Score} & 0.96 & 0.96 & 0.96 \\
\textbf{Variance} & $4.57\mathrm{e}{-5}$ & $4.89\mathrm{e}{-5}$ & $4.84\mathrm{e}{-5}$ \\
\bottomrule
\end{tabular}
\caption{Random forest classifier performance predicting Echo-Show5 button presses from collected network traffic. Average test scores and variances reported over 50 repetitions with randomly selected train/test set divisions.}
\label{tab:scores}
\end{table}      

These results also corroborate the possibility of inferring details of user interactions with consumer IoT devices from network traffic~\cite{apthorpe2019keeping, acar2020peek, trimananda2020packet} and show that our automated data collection approach could be useful for similar future studies. 
\section{Limitations}
\label{sec:limit}

Using a robotic arm to automate physical IoT device interactions is an effective way to scale the collection of IoT traffic across device behaviors. However, the nature of collecting network data from physical IoT devices via robotics has certain constraints:

\subsection{Non-Physical User Interface Elements}
While a robotic arm can perform many types of interactions with IoT devices, such as pressing buttons or sliding switches, it cannot perform non-physical interactions such as voice commands. Even certain physical user interface elements, such as touch screens, can prove challenging for off-the-shelf robotic arms. Unlike physical buttons, there are infinite possible interactions to test for comprehensive coverage of touch screen presses, and the location of touch screen input elements may change as the result of prior interactions. Considering that some IoT devices, including the Echo-Show5, have touch screens and voice commands as part of their major functions, it is important to explore additional techniques (such as software emulators) that can automate these types of interactions. 

Despite this limitation, we believe that the method described in this paper will be useful to IoT researchers. Physical device interfaces (e.g.,~buttons and switches) have proved the most challenging to automate thus far and are almost ubiquitous in inexpensive consumer IoT products.
Our proposed method is designed to facilitate the collection of network traffic on devices with such physical interfaces. Additionally, combining a robotic arm with a microphone and an instrumented smartphone running devices' associated mobile applications would be sufficient to explore all possible user inputs for many common devices without requiring
tedious manual button pressing or dedicated development of device-specific emulators. 

\subsection{IoT Device Size}
Configuring the robotic arm to interact with the Echo-Show5 and the Sensi-Thermostat was feasible as both devices are small enough for the robotic arm to reach any location on the devices. In general, the size of the robotic arm limits the size of the IoT device that can be tested. Significantly larger devices (such as a smart refrigerator) exceed the maximum reach or range of motion of most hobbyist robotic arms. While larger and more precise robotic arms are available on the market, they are substantially more expensive. One possible solution is to have multiple smaller robotic arms operate on a single large device simultaneously, but this would require precise coordination of the arms to interoperate. However, this limitation does not raise significant concern, since most popular consumer IoT devices are small appliances that would be suitable for our approach.

\subsection{Environmental Sensor Data}
Many IoT devices include environmental sensors, such as thermometers, light sensors, accelerometers, and gyroscopes that determine their behaviors and network communications~\cite{apthorpe2019keeping} in conjunction with user interface elements.
The scope of possible readings from these sensors is not explored by our robotic arm approach and would require a software emulator or a laboratory with local environment controls. While it would be possible to place both a robotic arm and an IoT device into a chamber with controllable temperature or lighting and conduct permutation-based testing with each of these variables, this is outside the scope of this paper. 

\section{Future Work}
\label{sec:future}

Apart from the limitations stated in Section~\ref{sec:limit}, our automated approach to IoT device interactions is amenable to many additions and improvements that could be the topic of future work. We hope that others will adopt and adapt this approach to automate and scale IoT research. 

\subsection{Randomness in Interaction Sequences}
One straightforward extension of our approach is to create permutation-based interaction sequences with increased randomness to better replicate the variety of real user interactions. Although our current implementation explores all possible unique interaction permutations, additional randomness could be added by randomly repeating button presses within interaction sequences or repeating entire interaction sequences for a random number of iterations.

A follow-up study could also test whether increases in interaction randomness actually increase the variety of collected network traffic. If network traffic is predominantly linked to the most recent interaction, introducing random delays in interaction frequency may not actually produce traffic of greater interest for follow-up analysis.

\subsection{Additional IoT Devices}
While we chose two popular consumer IoT devices to test with robotic arm interactions, we expect that this method is amenable to a wide variety of IoT products in consumer, medical, and other contexts. 
We are aware of several universities and consumer advocacy groups with access to many IoT devices and we recommend their use of our technique for automating network security and privacy analyses at scale. 
We would especially like to see this approach used to evaluate medical devices, as they often have many physical buttons and may be under-audited from a network security perspective~\cite{burns2016brief}. 

\subsection{Multi-Device Interactions}
Although this study explores automated interactions with individual devices, future work could apply our system to simultaneous interactions with multiple devices using multiple robotic arms. 
Some IoT devices are designed to communicate with other devices on the local network, and the traffic from these communications would not be visible from robotic arm interactions with a single device. With the help of a centralized coordinator (i.e., a parallel processing controller), a setup with multiple devices and multiple robotic arms could test interleaved interaction sequences across devices to record traffic from device-device communications. 
As discussed in~\cite{apthorpe2019keeping}, patterns of network traffic from multiple IoT devices within a household may allow privacy-violating inferences not achievable with traffic from a single device alone. 
Applying our approach to multiple devices could test the possibility of such inferences and potentially reveal other privacy or security vulnerabilities related to local device communications as well. 
As more manufacturers promote ``ecosystems'' of IoT devices, the ability to automate network research of multi-device interactions becomes increasingly important. 

\subsection{Computer Vision UI Identification}
The only manual process in our approach is to configure the robotic arm with the positions of each of the user interface elements on the device prior to automated interactions (Section~\ref{sec:inverse-kinematics}). This initialization step could also be automated in future work by adding a camera to the setup and using computer vision to identify the type and location of user interface elements on the device. This would provide an extremely low barrier to entry for network data collection from any IoT device at the expense of the additional camera hardware and increased potential for misidentification of UI elements. 

\subsection{Cyber-Physical Fuzz Testing}
Our approach could be used to fuzz test the user interfaces of IoT and other devices with essential physical interfaces, especially by entities such as security researchers or consumer protection groups without access to source code or device emulators. 
Testing all possible physical interactions with a device to see whether any produce buggy or malicious behavior could reveal previously unknown vulnerabilities. This idea could be combined with any of the other future directions in this section to expand the scope of automated security analysis for cyber-physical devices. 

\section{Conclusion}
\label{sec:conclusion}
The collection of network traffic data from consumer IoT devices has heretofore involved tedious manual interactions with devices in a laboratory setting or expensive crowdsourcing initiatives. In this paper, we present a novel method for automating the collection of IoT network traffic across a variety of user interactions and device behaviors: configuring a robotic arm to interact with physical user interface elements of IoT devices according to formalized permutation-based interaction sequences. 
We describe the steps required to implement this method, including applying inverse kinematics to obtain robotic arm movement parameters, physically enlarging IoT device buttons to improve interaction accuracy, and  
creating permutation-based interaction sequences to provide comprehensive coverage of possible user interface interactions. 

We apply this approach to two representative IoT devices, an Amazon Echo Show~5 and an Emerson Sensi Wi-Fi Smart Thermostat. 
We confirm its effectiveness through inspection of collected network traffic and by the successful creation of a machine learning model that can infer specific robot/device interactions from the collected traffic. 
This indicates that the collected data contains information about device behaviors that could be useful for network, security, or privacy analyses.

Compared to prior methods of IoT network data collection, our automated approach provides high interaction coverage and does not require tedious manual effort. Our method is readily adaptable to different consumer IoT devices and for testing multi-device interactions. 
We have made our source code and other reference materials needed for this approach publicly available for future research.



\ifCLASSOPTIONcaptionsoff
  \newpage
\fi


\bibliographystyle{IEEEtran}
\bibliography{main}

\end{document}